\documentstyle[aps,eqsecnum]{revtex}

\begin{document}
\draft
\title{Effects of motion in cavity QED} \author{A. C. Doherty, A. S.
  Parkins, S. M. Tan, D. F. Walls}
\address{Department of Physics, University of Auckland, Private Bag 92019, Auckland,\\
  New Zealand } \date{\today} \maketitle

\begin{abstract}
  We consider effects of motion in cavity quantum electrodynamics
  experiments where single cold atoms can now be observed inside the
  cavity for many Rabi cycles. We discuss the timescales involved in
  the problem and the need for good control of the atomic motion,
  particularly the heating due to exchange of excitation between the
  atom and the cavity, in order to realize nearly unitary dynamics of
  the internal atomic states and the cavity mode which is required for
  several schemes of current interest such as quantum computing.
  Using a simple model we establish ultimate effects of the external
  atomic degrees of freedom on the action of quantum gates. The
  perfomance of the gate is characterized by a measure based on the
  entanglement fidelity and the motional excitation caused by the
  action of the gate is calculated. We find
  that schemes which rely on adiabatic passage, and are not therefore
  critically dependent on laser pulse areas, are very much more robust
  against interaction with the external degrees of freedom of atoms in
  the quantum gate.
\end{abstract}


The realization of unitary dynamics of single quantum systems is a
field of great current interest. Firstly, emerging experimental
possibilities have led to the possibility of testing in the laboratory
many of the thought experiments regarding aspects of quantum mechanics
such as superposition, non-locality and entanglement which have
puzzled physicists since its inception \cite{haroche1998a}. Secondly a
new theoretical understanding of the possibilities of quantum
entanglement has led to interest in the efficacy of unitary quantum
evolution in problems of computation and communication \cite{pw1998a}.
In order to enforce unitary evolution for a given system it is
necessary to overcome any and all couplings to other degrees of
freedom. This is the problem of decoherence which typically leads to
increasingly classical seeming evolution as couplings to the
environment are increased. It is now the case however that several
experimental systems approach the idealized situation where a few
degrees of freedom are isolated almost entirely from their
surroundings with sufficient experimental control that almost
arbitrary unitary evolutions can be effected. These systems include
the motional and internal electronic states of ions in Paul traps
\cite{wineland1998a} and the system in which we are interested here,
single atoms in high finesse Fabry-Perot microcavities
\cite{berman1994a}.

Although experiments in cavity quantum electrodynamics (CQED) are far
from realizing full-blown quantum computers \cite{turchette1995b},
this system is particularly interesting since the environmental noises
and couplings to extraneous systems are particularly well understood.
Current experiments \cite{mabuchi1996a,hood1998a,mabuchi1998b} are
able to observe the interaction of single atoms with the mode of the
cavity over many Rabi cycles due to the use of magneto-optical
trapping and cooling of the atoms.  Current experimental efforts are
aimed at trapping the atom inside the cavity for essentially arbitrary
lengths of time. Obtaining complete control over the motion of the
atom while inside the cavity has proven to be a challenging
experimental task particularly in the face of very large heating rates
in current experiments. These are due to the repeated exchange of
excitation between the atom and the field, with the associated
momentum kicks to the atom. In this paper therefore we briefly discuss
effects of the motion of the atom in current and future experiments in
CQED with single cold atoms. In particular we will be interested in
the level of control of the motion which will be necessary in order to
have a system which corresponds to a quantum gate with a high level of
accuracy.

This paper is structured as follows. In section (\ref{sec1}) we
discuss the interaction of a single atom with a cavity mode and the
effects of motion on current experiments in cavity QED. We discuss the
reasons for seeking to confine the atom in a potential that is
independent of the cavity mode. In section (\ref{sec2}) we discuss two
schemes for quantum computation in cavity QED, set up a model to
include motional effects and a measure of how closely such a gate
system approaches the ideal unitary evolution. In section (\ref{sec3})
we discuss the effects of motion on a Raman scheme for quantum
computing and in section (\ref{sec4}) we give a parallel treatment for
a scheme based on adiabatic passage. In section (\ref{sec5}) we
conclude.

\section{Single Atom in a Cavity}

\label{sec1}

In the experiments of \cite{mabuchi1996a,hood1998a,mabuchi1998b} very
cold Cesium atoms are dropped into tiny single-mode Fabry-Perot
cavities.  The master equation for the system is well known. The
Hamiltonian for a two level atom interacting with a single driven mode
of the electromagnetic field in an optical cavity using the electric
dipole and rotating wave approximations (in the interaction picture
with respect to the driving laser frequency) is

\begin{eqnarray}
H &=&\frac{{\bf p}^{2}}{2m}+V({\bf r})+\hbar (\omega _{0}-\omega _{L})\sigma
_{+}\sigma _{-}+\hbar (\omega _{c}-\omega _{L})a^{\dagger }a+  \nonumber \\
&&\hbar g_{0}\psi ({\bf r})(a^{\dagger }\sigma _{-}+\sigma _{+}a)+\hbar
E(a^{\dagger }+a).  \label{hamiltonian}
\end{eqnarray}
The first term is the kinetic energy of the atom, the second describes
an external potential which in subsequent sections we will allow to
confine the atom in the absence of a cavity field. The next two terms
are the energy in the internal state of the atom and the cavity
excitation. The fifth term describes the position dependent
interaction of the cavity mode and the atomic dipole. The strength of
this interaction is determined by the single photon Rabi frequency
$g_{0}$ which depends on the cavity mode volume and the dipole matrix
elements for the relevant atomic transition. The final term describes
the driving of the cavity by a coherent (laser) driving field of
amplitude $E$, chosen here to be real.  The atomic transition
frequency is $\omega _{0}$, the cavity has a resonance at the
frequency $\omega _{c}$ and the driving frequency is $\omega _{L}$.
The cavity mode function is $\psi ({\bf r})=\cos (k_{L}x)\exp (-(y^{
  2}+z^{2})/w_{0}^{2})$, describing the Gaussian standing wave
structure of the field in the Fabry-Perot cavity, the optical
wavelength $\lambda _{L}=852.359$~nm for the Cesium transition
employed ($k_{L}=2\pi /\lambda _{L}$).

Dissipation in the system is due to cavity losses and spontaneous
emission.  By treating the modes external to the cavity as heat
reservoirs at zero temperature it is possible to derive the standard
master equation for the density operator of the system
\cite{carmichael1993a}, $\rho $,
\begin{equation}
\dot{\rho}=\frac{1}{i\hbar }[H,\rho ]+2\kappa  
{\cal D}[a]\rho +2\gamma {\cal D}[\tilde{\sigma}_{-}]\rho .
\label{mastereqn}
\end{equation}
The superoperator ${\cal D}[c]$ acting on a density matrix $\rho $ is ${\cal %
  D}[c]\rho =c\rho c^{\dagger }-\frac{1}{2}c^{\dagger }c\rho
-\frac{1}{2}\rho c^{\dagger }c.$ The dipole decay rate is $\gamma $,
while the cavity field decay constant is $\kappa $. The third term
describes the effect of spontaneous emission and $\tilde{\sigma}_{-}$
is an operator describing both the change of internal state and the
momentum kick on the atom due to a single spontaneous emission.

These experiments represent an important improvement on previous work
in that each atom remains in the cavity for many Rabi cycles. The
cooling of the atoms prior to entering the cavity effects a separation
of timescales of the dynamics of the external degrees of freedom of
the atom and the other degrees of freedom in the problem. The
variation of the coupling due to the atomic motion, frequency with
which the atom passes through wavelengths of the standing wave in the
cavity, is, at least initially, much less than the other frequencies
involved,
\begin{equation}
\left|{\bf p} \cdot \nabla \psi({\bf r})\right| \ll g_{0},\kappa ,\gamma .
\end{equation}
However the driving of the cavity field and the interaction of this
field with the atom leads to disturbance of the motion which can be
understood in terms of a semi-classical theory of the mechanical
effects of light in the cavity. In particular the effect of the dipole
force in trapping atoms near the antinodes of the cavity field was
observed in \cite{hood1998a}. However the rapid exchange of excitation
between the atom and the cavity mode leads to an increased momentum
diffusion or heating in this system over a free-space standing-wave.
The effect of heating was probably very significant in
\cite{mabuchi1996a,mabuchi1998b}. These semi-classical parameters can
be calculated numerically, as a function of atomic velocity, through a
matrix continued fraction calculation just as in the free-space
theory. Simulations of the classical trajectories of the atoms inside
the cavity can be performed in three dimensions using a Langevin
equation approach with the semiclassical force, friction and momentum
diffusion acting on a classical point particle
\cite{doherty1997a,mabuchi1998b}. It was found that the atom is in the
cavity long enough to be significantly heated and that only a few
atoms will be sufficiently slow that their motion along the standing
wave can be tracked. The heating of the motion means that the atoms
will eventually boil out of even the very deep potential wells that
can be set up by the dipole force due to the very large field
gradients inside the cavity.

One way to reduce the noise on the atom is to move into a highly
detuned regime, the dispersive limit of CQED, in which the atom
induces a phase shift on the field and the the dipole force provides a
nearly conservative potential for the atom. This corresponds to the
far off resonance trapping of atoms in optical lattices. In this limit
both the cavity field and the atomic internal state can be
adiabatically eliminated and a master equation written for the quantum
mechanical motional state alone \cite{doherty1998a}
\begin{mathletters}
\begin{eqnarray}
\dot{\rho} &=&\frac{1}{i\hbar }[H^{\prime },\rho ]+\frac{2g_{0}^{4}E^{2}}{%
\kappa ^{3}\Delta ^{2}}{\cal D}[\cos ^{2}(k_{L}x)]\rho \\
H^{\prime } &=&\frac{p_{x}^{2}}{2m}-\hbar \frac{g_{0}^{2}E^{2}}{\kappa
^{2}\Delta }\cos ^{2}(k_{L}x).
\end{eqnarray}
The Lindblad term describes heating due to light scattering caused by
cavity assisted spontaneous emission. Essentially this is an extra
contribution to the light scattering heating present in far off
resonant optical lattices and takes place even though the atom is in
principle never excited. Such a regime does provide the hope of long
trapping times but there are technical difficulties associated with
attaining sufficiently high detunings to fully realize the model,
although effects such as light scattering due to free space
spontaneous emission could easily be included in this treatment.

There are several reasons to have some other means of trapping the
atom in the cavity. Firstly even in the far detuned regime driving the
cavity field does not give a particularly slow heating environment for
the atom due to the increased light scattering out through the cavity
mirrors. Secondly, quantum computing and other interesting schemes for
this atom cavity system tend to require that the cavity field is
initially in the vacuum state and is not driven and that the motion of
the atom is undisturbed during the action of the gate. So it is
enticing to consider loading the cavity and trapping the atom with an
optical lattice, perhaps another --- lower finesse --- mode of the
Fabry-Perot, or with an ion trap. We anticipate that in any practical
realization of these models there will be a means of confining the
atom other than the cavity field alone.

It is important that the heating rate $\gamma _{\text{heat}}$ of the
atom in this potential be slow compared to the other dynamics of the
system and we anticipate that as in the dispersive regime considered
above the system will preserve the situation present initially in
current experiments, that the frequency $\omega $ associated with the
atomic motion is small
compared to $%
\kappa ,\gamma $ which are in turn small compared to the coherent
coupling $%
g.$ \ Conditions which will ultimately realize unitary evolution of
the atomic internal states and the cavity field will therefore be,
\end{mathletters}
\begin{eqnarray*}
g &\gg &\kappa ,\gamma , \\
g &\gg &\omega \gg \gamma _{\text{heat}}.
\end{eqnarray*}
As we will see below there is also a regime in which the mechanical
motion of the atom is much faster than the internal state evolution
and therefore effectively decouples from it. However the heating of
the atom would still have to be negligible for at least a few Rabi
cycles so that a gate could be performed before the motional state had
to be reset. This requires
\[
\omega \gg g\gg \gamma _{\text{heat}}
\]
which implies very much larger quality factors than current or near
future optical lattice or ion trap technology could provide. Finally
the requirement that the action of the gate not affect the motion of
the atom too greatly will mean that transitions between harmonic
oscillator states require more energy than is provided by the atom
emitting or absorbing a photon. Thus we also require that the recoil
frequency $\omega _{r}=\hbar k_{L}^{2}/2m$ for the atom and transition
under consideration is small compared to the motional frequency
\[
\omega \gg \omega _{r}.
\]

\section{Quantum Computing in CQED}

\label{sec2}

There are several proposals for realizing quantum gates in CQED with point
dipoles. We wish to consider here the unavoidable effects of the motion and
thus understand the level of control of the motion which will be
necessary 
to realize these schemes with a high fidelity. In particular we wish to
compare a system based on controlling the times for which a given
interaction is turned on and off with one which relies on an adiabatic
passage through eigenstates of a time-dependent Hamiltonian and for which
the interaction time is not critical.

\subsection{Raman Scheme}

A model of the first type is given by van Enk {\em et al}
\cite{vanenk1997a} which employs a Raman transition in a cavity to
effect a two bit quantum gate. The procedure obtains conditional
dynamics for the two atomic internal states through the exchange of a
cavity photon between the two atoms, which are imagined to be confined
to the antinode of the cavity field. It is also assumed the atoms can
each be driven through the side of the cavity by a separate laser.
Each atom has two states $|0\rangle _{i},|1\rangle _{i}$which form the
qubit, an auxillary level $|r\rangle _{i}$ which is coupled to the
cavity and an excited
state $%
|e\rangle _{i}$ from which the laser driving is detuned. A Raman
transition is employed since this reduces the effect of spontaneous
emission.  The interaction Hamiltonian for the atom cavity interaction
with the excited state adiabatically eliminated is
\begin{equation}
H=\sum_{j=1,2}\frac{gf_{j}(t)}{2}|1\rangle _{jj}\langle r|a+\text{H.c.}
\end{equation}
where the atoms have the level structure shown in figure
(\ref{level}). The function $\ f_{i}(t)<1$ describes some laser
driving pulse shape and the constant $g$ describes the effective Raman
coupling. Since both atoms interact with the cavity mode this
Hamiltonian can be used to build a quantum gate. This gate can be
designed so that population in $|0\rangle $ rarely has to interact
with the cavity thereby reducing the errors. In \cite {vanenk1997a} a
sequence of pulses is described which realizes the universal two-bit
gate
\begin{eqnarray*}
|0\rangle _{1}|0\rangle _{2} &\rightarrow &|0\rangle _{1}|0\rangle
_{2};\;|1\rangle _{1}|0\rangle _{2}\rightarrow -|1\rangle
_{1}|0\rangle _{2} 
\\
|0\rangle _{1}|1\rangle _{2} &\rightarrow &|0\rangle _{1}|1\rangle
_{2};\;|1\rangle _{1}|1\rangle _{2}\rightarrow |1\rangle _{1}|1\rangle
_{2}. 
\end{eqnarray*}

\subsection{Adiabatic Passage via Dark State }

A model of the second type is given by Pellizzari {\em et al} \cite
{pellizzari1995a} who show how to perform a controlled-NOT\ and
various other quantum gates by encoding two qubits onto four levels of
a single atom and employing laser driving to achieve the conditional
dynamics. Information about one of the qubits is transferred back and
forth between the two atoms in the gate by transferring coherences
between the ground states of one atom to the other through an
adiabatic passage involving excitation of the cavity field. This
adiabatic passage is through a dark state based on the single atom
dark states discussed in \cite{parkins1993a} and thus suppresses
spontaneous emission without employing a Raman transition since the
excited states of the atoms are in principle never occupied. The
interaction Hamiltonian is very similar to the one for the previous
system
\begin{equation}
H=\sum_{j=1,2}\frac{g}{2}|e\rangle _{jj}\langle r|A+\frac{\Omega
  _{j}(t)}{2} |e\rangle _{jj}\langle 1|A+\text{H.c.}
\end{equation}
and this Hamiltonian has the dark states
\begin{eqnarray*}
|D_{0}\rangle  &=&|r,r,0\rangle \equiv |r\rangle _{1}|r\rangle
_{2}|0\rangle 
_{c} \\
|D_{1}\rangle  &\propto &\Omega _{1}g|r,1,0\rangle +\Omega
_{2}g|1,r,0\rangle -\Omega _{1}\Omega _{2}|r,r,1\rangle .
\end{eqnarray*}
Here we have labeled the states in the same way as in the previous
model although $|r\rangle $ could in fact be used as the logical zero
of the qubit. With the second atom initially prepared in $|r\rangle$,
switching on the laser driving the second atom, so that $\Omega_{2}$
is initially large, and then slowly increasing the driving on the
first atom while decreasing the driving on the second, transfers the
state of the
first atom on $%
\{|r\rangle ,|1\rangle \}$ to the second atom. With appropriate
driving on the transition $|r\rangle \leftrightarrow |0\rangle$ the
logical state of the first atom is transfered to the logical state of
the second atom. In order to perform a gate it is necessary to
consider more than these three levels on atoms and the two qubits can
be transferred to coherences between four ground states of the second
atom on which the gate can be performed through Raman transitions
between the ground states. Since this laser driving can be achieved in
such a way that all of the field gradients in the vicinity of the atom
will be small we can disregard motional effects for this part of the
evolution. What is of interest is the motional state dependence of the
actual adiabatic transfer of coherence through the cavity field
described here so we will restrict ourselves to the simpler three
level system and investigate the behavior of the dark state when
motional states and the position dependence of the coupling to the
cavity are included.

\subsection{Errors in Quantum Computers}

Several strategies for overcoming the effects of unwanted couplings to
the environment --- including the motion of the atom --- are possible.
One approach, that of quantum error correcting codes, first described
by Shor and Steane \cite{shor1995a,steane1996a}, and encoded gate
operations, follows from the realization that errors during the
storage of a quantum state or its manipulation can be corrected by
coding the qubits in larger Hilbert spaces made up of several
identical quantum systems. This would correspond in our case to
several atoms in a single cavity or perhaps several cavities each with
their own atoms in order to achieve the redundancy necessary to
overcome the effects of environmental noise. In this case it can be
shown that given sufficient resources any quantum computation can be
performed as long as the fundamental error rate for each of the
systems is below some threshold value \cite{knill1998a}. If this
approach is taken the critical thing to know for a particular
candidate system is the fundamental error rate due to a given coupling
to the environment and what can be done to modify this rate. Other
strategies are system specific and revolve around characterizing the
errors that occur in a given physical realization of a quantum
computer and correcting for those errors alone, perhaps to all orders.
Work along these lines is very far advanced in the case of CQED
quantum computing, \cite{vanenk1997a} and references therein.

In the work of van Enk {\em et al} \cite{vanenk1997a} it i s necessary
that a stationary property holds for the coupling to the environment.
This essentially requires that the evolution operators which entangle
the basis vectors of the computational subspace with the environment
depend only on the length of time it takes to operate the gate and
that they commute with each other. This allows schemes which
symmetrize the effect of the noise on all of the basis states of the
qubits by allowing them to interact with the environment at different
times. Studying the effects of motion on the CQED quantum computer is
important because, unlike photon absorption, spontaneous emission and
systematic errors in the driving times, errors induced by the motion
do not obey this stationary property. This is because the free
evolution term and the position dependent coupling term do not
commute. In any situation where the motion of the atom is comparable
to the coupling $g,$ so that the evolution operators associated with
these terms in the Hamiltonian will be significantly non-commuting,
there will be a departure from the conditions required for the
correction schemes of \cite{vanenk1997a} to work ideally. Therefore
the rate of errors due to the motion will be one contribution to the
noise processes which limit the performance of the CQED quantum
computer with currently available simplified approaches to error
correction.

\subsection{Model for Motion of Atoms}

We need a model to describe the fundamental effects of motion in CQED
models such as those discussed above. The model we will use is
motivated by the differences of timescales discussed in the previous
section and the knowledge of correction schemes for photon losses and
spontaneous emission to all orders. The important rates are therefore
the coherent coupling rate achieved for the gate, the frequency
associated with the motion of atom and the recoil frequency which
describes the effect of the emission or absorption of a single photon
on the motion of the atom. Position dependence of the coupling
combined with the initial spread of the motional state and motion of
the atom in the trapping potential will be sources of noise in the
computation. We will assume that the particle is trapped in a harmonic
potential with a low heating rate. If the atom is cold and confined in
a far off resonance dipole trap then this will be the most significant
contribution to the effects of the motion although the oscillation
frequency may depend on the internal state of the atom. If the
oscillation rate in the potential is small compared to the effective
coupling frequency $g$ then in effect we are just adjusting the length
scale of the initial state of motion and the precise shape of the
potential and any dependence of the potential on the internal state
will not significantly change the results.  We will leave the
consideration of heating of the atom during the gate action for future
work although in the limit in which the motion is rather slower than
the coupling and the heating rate is slower again than this then the
only significant effects of heating will be to lead to initially
thermal states of the motion which we will consider in the following.

Assuming then that each atom is trapped in a standing wave cavity in a
harmonic potential that is the same regardless of the internal state
of the atom we have the Hamiltonian
\begin{equation}
H_{i}=\frac{g_{i}}{2}\cos \left( \eta \left( A_{i}+A_{i}^{\dagger }\right)
\right) \left( \tilde{\sigma}_{i}+\tilde{\sigma}_{i}^{\dagger }\right)
+\omega A_{i}^{\dagger }A_{i},  \label{motqg}
\end{equation}
where we have defined the operator $\tilde{\sigma}_{i}=|1\rangle
_{ii}\langle r|a$ and $A_{i}$ is the lowering operator for the
motional state of the atom. The recoil frequency $\omega _{r}=\hbar
k_{L}^{2}/2m$ and
the oscillation frequency $\omega $ define a Lamb-Dicke parameter $\eta =$ $%
\sqrt{\omega _{r}/\omega }.$ In the ideal situation the atom is
tightly confined compared to the wavelength of the light and this
Lamb-Dicke parameter is very small. We assume that the couplings to
the lasers driving the atoms are position independent, this makes
sense because these lasers reach the atom through the sides of the
cavity and the beam width will in general be much larger than a
wavelength. This then is the model depicted schematically in figure
(\ref{schem}). There will in fact be some interaction between errors
caused by the motion and by cavity decay, which is the most important
feature left out of our model, however the purpose of this work is to
identify the ultimate sources of error due to the motion in the
situation in which the effects of cavity decay can in principle be
reversed to all orders as in \cite{vanenk1997a}.

\subsection{Characterising Imperfect Gates}

We also need a means of characterizing the success and failure of a
gate. In essence we need a measure of the distance between the actual
evolution and the ideal one. Several such measures of the distance
between superoperators have been used or proposed in related work
\cite {poyatos1997a,giedke1998a,aharanov1998a}. In this case we are
just interested in a simple measure which is physically motivated for
quantum gates. We will employ a simple modification the entanglement
fidelity introduced in \cite{schumacher1996a}. This is related to the
overlap or
fidelity of a state $\rho $ to some desired pure state $|\psi \rangle$, $%
F=\langle \psi |\rho |\psi \rangle$. $F$ is one if and only if $\rho
=|\psi
\rangle \langle \psi |$. The entanglement fidelity for a noisy evolution $%
{\cal E}$ on some state $\rho$ of a system $Q$ is
\begin{equation}
F_{e}(\rho ,{\cal E})=\langle \psi ^{RQ}|\left( {\cal I}^{R}\otimes {\cal E}%
\right) \left( |\psi ^{RQ}\rangle \langle \psi ^{RQ}|\right) |\psi
^{RQ}\rangle 
\end{equation}
where $|\psi ^{RQ}\rangle $ is a pure state of $Q$ and a fictional
auxiliary system $R$ such that Tr$_{R}(|\psi ^{RQ}\rangle \langle \psi
^{RQ}|)=\rho $ and ${\cal I}^{R}$ is the identity superoperator on
$R.$ It is shown in \cite{schumacher1996a} that $F_{e}$ is independent
of the particular purification $|\psi ^{RQ}\rangle $ chosen. The
entanglement fidelity can be thought of as
characterizing how well the state and its entanglement are preserved by $%
{\cal E}$. It is shown in \cite{nielsen1996a} that
\[
F_{e}\left( \rho ,{\cal E}\right) =\min_{\rho ^{RQ},{\cal E}^{\prime
    }}F\left( \left( {\cal E}^{\prime }\otimes {\cal I}^{Q}\right)
  (\rho ^{RQ}),\left( {\cal E}^{\prime }\otimes {\cal E}\right) \left(
    \rho ^{RQ}\right) \right)
\]
where $F$ is the fidelity of mixed states defined in \cite{jozsa1994a}
and describes how close two density matrices are to each other. $\rho
^{RQ}$ is
an extension of the state $\rho $ to the combined system such that Tr$%
_{R}(\rho ^{RQ})=\rho $ and ${\cal E}^{\prime }$ is an arbitrary
evolution on the auxiliary space $R$. Thus the entanglement fidelity
corresponds to the worst possible fidelity of the system state after
the evolution ${\cal E} $ to its initial state regardless of how the
system is entangled with the environment and of what dynamics ${\cal
  E}^{\prime }$ the environment is undergoing. The entanglement
fidelity provides a good measure of the preservation of a state in the
memory of a quantum computer which could be entangled with many other
qubits in the computer and where these qubits could be undergoing
arbitrary evolutions as part of the computation.  Moreover if $\rho
=\sum p_{i}|\psi _{i}\rangle \langle \psi _{i}|$ then the
entanglement fidelity is less than or equal to the average fidelity under $%
{\cal E}$ of the ensemble making up $\rho ,$ $F_{e}\leq \sum
p_{i}\langle \psi _{i}|{\cal E}\left( |\psi _{i}\rangle \langle \psi
  _{i}|\right) |\psi _{i}\rangle .$ But on the other hand if the
fidelity of all of the pure states $|\psi \rangle $ with support on
$\rho $ is close to one then the entanglement fidelity is close to one
also \cite{knill1997a}.

Motivated by these considerations we will use a gate entanglement
fidelity which measures how close ${\cal E}$ is to the ideal unitary
evolution $U$ over the whole computational subspace of $\{|0\rangle
_{1}|1\rangle _{1}|0\rangle _{2}|1\rangle _{2}\}$ by
\begin{equation}
F_{eg}\left( {\cal E},U\right) =\langle \psi ^{RQ}|U^{\dagger }\left( {\cal I%
}^{R}\otimes {\cal E}\right) \left( |\psi ^{RQ}\rangle \langle \psi
^{RQ}|\right) U|\psi ^{RQ}\rangle 
\end{equation}
where the $\rho $ is the completely mixed state on the computational
subspace Tr$_{R}(|\psi ^{RQ}\rangle \langle \psi ^{RQ}|)=\rho ={\cal I}%
^{C}/4.$ Thus if $F_{eg}$ is close to one then the gate is close to
ideal for all initial states of the two qubits regardless of how they
are entangled with the other qubits in computer and of how these other
qubits are being manipulated during the gate operation. This measure
has the property of measuring not just how close the evolution is to
the ideal evolution for any pure state on the computational subspace
but also how well the evolution preserves entanglement between the
state of the system and the state of \ other systems which may be part
of the quantum computer.

\section{Gate Fidelity for Raman Scheme}

\label{sec3}

Position dependence of the coupling and motion of the atom in the
trapping potential will be a source of noise.  In order that the atom
in fact be localized near the antinode of the cavity it should occupy
a motional state of low excitation in the potential and have a recoil
frequency $\omega _{r}=\hbar k_{L}^{2}/2m$ much smaller than the
oscillation frequency of the atom. Assuming that this is the case we
can perform time-dependent perturbation theory in order to find the
fidelity of the fundamental entangling evolution. If we do this to
$O(\omega _{r}^{2}/\omega ^{2})\,$we can consider the simplified
Hamiltonian
\begin{mathletters}
\begin{eqnarray}
H &=&H_{0}+V, \\
H_{0} &=&\frac{g}{2}\left( \tilde{\sigma}+\tilde{\sigma}^{\dagger }\right)
+\omega A^{\dagger }A, \\
V &=&-\frac{\eta ^{2}g}{4}\left( A+A^{\dagger }\right) ^{2}\left( \tilde{%
\sigma}+\tilde{\sigma}^{\dagger }\right) 
\end{eqnarray}
\end{mathletters}
By moving into the interaction picture defined by $H_{0}$ it is
possible to do time-dependent perturbation theory to calculate
approximate states for the system and therefore the entanglement
fidelities and motional excitation. If we restrict our interest just
to one atom in the cavity for the moment then we can calculate the
Schr\"{o}dinger picture ket where the internal and cavity states are
initially $|1\rangle _{1}|0\rangle _{\text{c}}$ and then leave the
laser on such that in the idealized (point-dipole) case we end in the
state $|r\rangle _{1}|1\rangle _{\text{c}}$. We wish to leave open the
possibility of tailoring the length of the laser pulse such that the
fidelity of the final state is optimized by using a pi-pulse
appropriate to the mean-squared position of the atom. Thus we choose
the interaction
time $%
t=\pi (1+\delta )/g$ where $\delta$ will be of order $\omega
_{r}/\omega $ and will be chosen to maximize the fidelity of the final
state. We will consider initially just a number state of the atom and
perform the thermal average at the end of the calculation.

The overall Schr\"{o}dinger picture state after the evolution is
\begin{eqnarray*}
&&\left( 
\begin{array}{c}
1-\frac{1}{2}\left( \frac{\pi \eta ^{2}}{4}\left( 2n+1\right) -\frac{\pi
\delta }{2}\right) ^{2} \\ 
-\left( \frac{\eta ^{2}g}{4\omega }\sin (\pi \omega /g)\right) ^{2}\left(
n^{2}+n+1\right) 
\end{array}
\right) |r\rangle _{1}|1\rangle _{\text{c}}|n\rangle _{\text{m}1} \\
&&+i\left( \frac{\pi \eta ^{2}}{4}\left( 2n+1\right) -\frac{\pi \delta }{2}%
\right) |1\rangle _{1}|0\rangle _{\text{c}}|n\rangle _{\text{m}1} \\
&&+\left( \frac{\eta ^{2}g}{8\omega }\sqrt{(n+1)(n+2)}\left( 1-e^{-2i\pi
\omega /g}\right) \right) |1\rangle _{1}|0\rangle
_{\text{c}}|n+2\rangle _{\text{m}1} \\ 
&&-\left( \frac{\eta ^{2}g}{8\omega }\sqrt{n(n-1)}\left( 1-e^{2i\pi \omega
/g}\right) \right) |1\rangle _{1}|0\rangle _{\text{c}}|n-2\rangle
_{\text{m}1}. 
\end{eqnarray*}
where we have only retained those terms which turn out to affect the
fidelity and entropy up to fourth order in the Lamb-Dicke parameter
and have disregarded an overall phase.  Clearly the majority of the
population is in the desired final state, there is also population
left in the original internal state and a superposition of motional
states.

We assume that the initial motional state is in fact a thermal state
of average excitation $\bar{n}.$ The thermal averaging can be
performed by summing the series for the terms in the reduced density
matrix of the internal and cavity states resulting from each of the
individual initial number states since these are just geometric series
or their derivatives. The fidelity for this interaction with the
cavity is
\begin{eqnarray}
F &=&1-\frac{\pi ^{2}\delta ^{2}}{4}+\frac{\pi ^{2}\eta \delta }{4}%
\left( 2\bar{n}+1\right) -\left( \frac{\pi \eta ^{2}}{4}\right)
^{2}\left( 8\bar{n}^{2}+8\bar{n}+1\right)   \nonumber \\
&&-\left( \frac{\eta ^{2}g}{2\omega }\right) ^{2}\sin ^{2}(\pi \omega
/g)\left( \bar{n}^{2}+\bar{n}+\frac{1}{2}\right) .
\end{eqnarray}
We may have sufficient control over the length of the laser pulse that
we can choose $\delta $ so as to maximize this quantity, thus giving
us the best possible fidelity of the evolution. Setting $\delta =\eta
^{2}\left( 2\bar{n}+1\right) /2$ gives us
\begin{eqnarray}
F_{\text{opt}} &=&1-\left( \frac{\pi \eta ^{2}}{2}\right) ^{2}\left
  ( \bar{n}^{2}+\bar{n}\right)   \nonumber \\
&&-\left( \frac{\eta ^{2}g}{2\omega }\right) ^{2}\sin ^{2}(\pi \omega
/g)\left( \bar{n}^{2}+\bar{n}+\frac{1}{2}\right) .
\end{eqnarray}

These expressions show the basic behavior of the system in a number of
regimes. In the most relevant limit that the harmonic oscillation is
much
slower than the internal dynamics $\omega \ll g$ we get $F_{\text{opt}%
}\simeq 1-2\left( \pi \eta ^{2}/2\right) ^{2}\left( \bar{n}^{2}+\bar{n}+%
\frac{1}{4}\right)$. In this case the motion of the atom is irrelevant
during the time for a pi-pulse and so in this case the parameters
$\eta $ and $\bar{n}$ essentially just define the initial position
spread and coherence length of the atomic motional state. This is the
fidelity that would be achieved for any such initial state regardless
of the details of the atomic motion on longer timescales. The opposite
limit of very fast motion $g\ll \omega \,\ $amounts to a rotating wave
approximation for the mechanical motion in which the atom oscillates
in the potential many times during a single operation,
$F_{\text{opt}}\simeq 1-\left( \pi \eta ^{2}/2\right) ^{2}\left(
  \bar{n}^{2}+\bar{n}\right) .$ This limit is attractive since it
suggests that if the oscillator is sufficiently cold then the effects
of motion could be overcome simply by modifying the naive length for a
pi-pulse of the system. However this still requires the assumption
that the heating rate of the motion $\omega _{\text{heat}}\ll g$,
which implies an enormous quality factor for the mechanical motion.
The laser power required to achieve $g\ll \omega$ in a far off
resonant optical trap would probably be prohibitive in any case. Noise
in current ion trap experiments would result in heating rates that
were at least comparable with the couplings $g$ so such a regime would
appear to be unfeasible with near future technology.

Neither will the computing operations leave the motional state
unmodified.  The state will be heated until eventually it will become
necessary to cool the motion of the atom. An indication of this can be
found by calculating the excitation of the motional state after one
exchange on excitation between the atom and the cavity. This is
entirely due to contributions resulting from transitions between
motional states at some stage during the evolution and as such depends
on trigonometric functions of the ratio between coupling and
mechanical oscillation frequencies and is independent of small changes
in the length of the laser driving,
\begin{equation}
\langle A^{\dagger }A\rangle -\bar{n}=\left( \frac{\eta ^{2}g}{2\omega }%
\right) ^{2}\sin ^{2}(\pi \omega /g)\left( 2\bar{n}+1\right) .
\end{equation}
So that in the rotating wave regime $g\ll \omega $ the effective
decoupling of the internal state and cavity dynamics from the motion
means that the motional state is unaffected by the action of the gate.
On the other hand in the more realistic situation $\omega \ll g,$
$\langle A^{\dagger} A\rangle -\bar{n}\simeq \left( \pi \eta
  ^{2}/2\right) ^{2}\left ( 2\bar{n}+1\right)$.

It is straightforward to extend this calculation to the full evolution
of the quantum gate with two atoms in the cavity described above and
to evaluate the entanglement fidelity for the gate operation
\begin{eqnarray}
\label{entfid}
F_{eg} &=&1-\pi ^{2}\eta ^{4}\left( \bar{n}^{2}+\bar{n}+\frac{1}{8}\right) 
\\
&&-\frac{\eta ^{2}g^{2}}{4\omega }\sin ^{2}(\pi \omega /g)\left( 1+\cos
^{2}(\pi \omega /g)\right) \left( \bar{n}^{2}+\bar{n}+\frac{1}{2}\right)  
\nonumber
\end{eqnarray}
which we give here for the situation in which the driving is not
optimized.  In this limit of fidelity close to one the entanglement
fidelity essentially reduces the to average of the fidelities of the
gate operation on each of the four basis states of the computational
subspace, although this is not true in general.

The motion of the atom will be excited by the action of the gate
depending on the actual initial state of the gate. However in the
operation of the gate this initial state could be any superposition of
these and could be entangled with the state of other qubits in the
computer. As a
measure of the overall heating of the motion we will calculate $n_{i}=$Tr$%
\left( A_{i}^{\dagger }A_{i}{\cal E}\left( |\psi ^{RQ}\rangle \langle
    \psi ^{RQ}|\right) \right) $ for a purification on the
computational
subspace of $%
\rho ^{C}={\cal I}^{C}/4$ as discussed above. This basically assumes
no knowledge of the internal state and therefore averages the effect
of the motion for each of the four basis states of the computational
subspace.  These excitation parameters can be calculated
\begin{mathletters}
\begin{eqnarray}
n_{1}-\bar{n} &=&\left( \frac{\eta ^{2}g}{2\omega }\right) ^{2}\sin
^{2}(\pi 
\omega /g)\left( \allowbreak 2\bar{n}+1\right),  \label{anheat} \\
n_{2}-\bar{n} &=&\left( \frac{\eta ^{2}g}{4\omega }\right) ^{2}\sin
^{2}(2\pi \omega /g)\left( \allowbreak 2\bar{n}+1\right). 
\end{eqnarray}
The different dependence on $\omega /g$ is due to the motional states
of the atoms being excited at different times during the action of the
gate.

We have also performed numerical simulations of the Hamiltonian
(\ref{motqg} ) with laser pulses as described above to all orders in
the Lamb-Dicke parameter, employing a number state expansion of the
motional operators.  Here we will plot results for initial states
where both the cavity and the atomic motion are initially in the
ground states $|\psi ^{RQ}\rangle =|\psi
^{RC}\rangle |0\rangle _{c}|0\rangle _{1m}|0\rangle _{2m}$ where Tr$%
_{C}\left( |\psi ^{RC}\rangle \langle \psi ^{RC}|\right) ={\cal
  I}^{C}/4.$ In figure (\ref{entfid1}) the gate entanglement fidelity
for this procedure is plotted as a function of the Lamb-Dicke
parameter along with the analytic approximation resulting from
equation (\ref{entfid}). This approximation is seen to hold up to
reasonably large values of $\eta .$ We considered recoil frequencies
sufficiently small that the entanglement fidelity and motional
excitation were effectively independent of $\omega $ except through
$\eta $, the actual values used in these simulations were $g=1,\omega
_{r}=0.0005$.  Once $\eta \simeq 0.4,$ errors are in the region of
$10\%$. The motional excitation of the first atom is plotted in figure
(\ref{heat1}) along with the approximation of equation (\ref{anheat}).

\section{Adiabatic Passage and Motion} \label{sec4}

A comparison of the previous scheme with one involving adiabatic
passage is motivated by the fact that adiabatic passage schemes do not
depend on the pulse area of the laser pulses. The position variation
of the coupling means that different parts of the wave function see
different pulse areas and so may be under or over-rotated by the
driving laser. All that is required for the adiabatic theorem to hold
is that the Hamiltonian is varied sufficiently slowly that
non-adiabatic transitions, the rate of which depend on the energy
separation between the eigenstates, do not occur, see for example
\cite{messiah1962a}. The energy spacing of the eigenstates from
neighboring eigenstates is determined by the size of the coupling $g$
so it might be hoped that it would be practical to perform the
transfer sufficiently slowly that all the population in a wide area
around the antinode of the field was transferred through the dark
state with high fidelity.

Another way of seeing this is to consider the eigenstates of the
Hamiltonian with motion included. Dark states exist which can effect
the transfer as long as $\omega \ll g$ --- a kind of Raman-Nath regime
for the gate. The values $i,j,k,l,m$ in the ket $|i,j,k,l,m\rangle
\equiv |i\rangle_{1} |j\rangle_{2} |k\rangle_{\text{c}}
|l\rangle_{\text{m}1} |m\rangle_{\text{m}2}$ refer to the internal
state of the first atom, of the second atom, the cavity state, the
motional state of the first atom and of the second atom, respectively.
The states
\end{mathletters}
\begin{eqnarray*}
|D_{1}\rangle  &\propto &\Omega _{1}g|r,1,0,n_{1},n_{2}\rangle +\Omega
_{2}g|1,r,0,n_{1},n_{1}\rangle  \\
&&-\Omega _{1}\Omega _{2}|r,r,1,n_{1},n_{2}\rangle  \\
&&-\eta ^{2}\Omega _{1}\Omega _{2}\left( A_{1}^{\dagger }+A_{1}\right)
^{2}|r,r,1,n_{1},n_{2}\rangle  \\
&&-\eta ^{2}\Omega _{1}\Omega _{2}\left( A_{2}^{\dagger }+A_{2}\right)
^{2}|r,r,1,n_{1},n_{2}\rangle 
\end{eqnarray*}
are eigenstates of the Hamiltonian including atomic motion up to
$O(\eta ^{4})$. Thus terms in the Hamiltonian of $O(\eta ^{2})$ do not
cause errors in the adiabatic passage as they do in the Raman scheme.
As a result we could expect fidelities for the process differing from
one by numbers of $O(\eta ^{8}).$

We simulated the adiabatic transfer of coherence discussed above for $%
g=1,\omega _{r}=0.0001$. We used Gaussian pulse profiles
$f_{i}(t)=\exp \left( -\left( g\left( t-t_{i}\right) /40\right)
  ^{2}/2\right) $ where the two pulses were separated by a time
$\Delta t=t_{2}-t_{1}=80/g.$ This choice resulted in transfer with
high fidelity and little population of the atomic excited states for
the point dipole atom. Figure (\ref{fid2}) plots the fidelity of the
transfer for different values of $\eta .$ Any gate built on this
principle will be limited by the fidelity of this procedure. Laser
pulses into the side of the cavity will have very much less affect on
the motion and will be achievable with high fidelity presuming the
atoms can be addressed separately by the lasers. Thus we do not show a
full gate entanglement fidelity for the gate described in
\cite{pellizzari1995a} here but it will be of the same order as the
fidelity for the transfer if laser operations are performed
accurately. The striking feature of figure (\ref {fid2}) is that for a
much larger range of $\eta $ the fidelity is essentially undisturbed
by the motion. As $\eta $ is increased the transfer takes place with
increased cavity excitation and motional excitation, but without any
increase in the excited state population, as suggested by the
approximate dark state above. Thus for $\eta \simeq 0.4$ errors are
still only 0.2\% although the fidelity has begun a sharp decline.
Simulation of the internal and external degrees of freedom for two
atoms as well as the cavity mode requires a very large Hilbert space
which is computationally intensive and so we have not explored values
of $\eta $ beyond those plotted here. So that the point at which the
fidelity becomes unusefully small is yet to be established.

\section{Conclusions}  \label{sec5}

We have investigated the effect of motion on experiments in cavity
QED. In particular we discussed the necessity of good control over the
motional state in order to realize the dynamics expected for models
involving point dipole atoms. The ultimate limitations that the
motional state places on quantum computing in CQED systems was
discussed for both a Raman scheme and one involving adiabatic passage
via a dark state to transfer information between the atoms. The scheme
involving adiabatic passage was found to be extremely robust to the
precise nature of the atomic motion which may be an important
consideration in future experimental implementations of similar
schemes.

\section*{Acknowledgments}

This work is supported by the Marsden Fund of the Royal Society of New
Zealand.

\begin{figure}[tbp]
\caption{Level structure of the atoms in cavity QED quantum computing
models. Information is typically encoded on the states $\{|0\rangle
,|1\rangle \}$.}
\label{level}
\end{figure}

\begin{figure}[tbp]
\caption{Schematic of imagined CQED quantum gate. Two laser beams drive
atoms which are harmonically trapped at antinodes of a high finesse
microcavity}
\label{schem}
\end{figure}

\begin{figure}[tbp]
\caption{Entanglement fidelity for quantum gate with motion as a
  function of the Lamb-Dicke parameter $\protect\eta $. Both atoms are
initially in the ground states of their motion. The solid line is from
numerical
calculations to all orders of $\protect\eta $ while the dotted line
represents an analytical approximation up to $O(\protect\eta ^{4}).$}
\label{entfid1}
\end{figure}
\begin{figure}[tbp]
\caption{Graph of average excitation $\langle a^{\dagger }a\rangle $ of one
  atom after the action of the quantum gate as a function of the
  Lamb-Dicke parameter $\protect\eta .$ The atoms are initially in the
  ground state of their motion. The solid line represents the results
  of numerical computations to all orders of $\protect\eta $ while the
  dotted line represents an analytical approximation to
  $O(\protect\eta ^{4}).$}
\label{heat1}
\end{figure}
\begin{figure}[tbp]
\caption{Fidelity of the adiabatic transfer of from one atom to another as
a function of the Lamb-Dicke parameter with atoms initially in the
ground state of their motion. Note the improved performance
compared to figure (\ref{entfid1}).}
\label{fid2}
\end{figure}

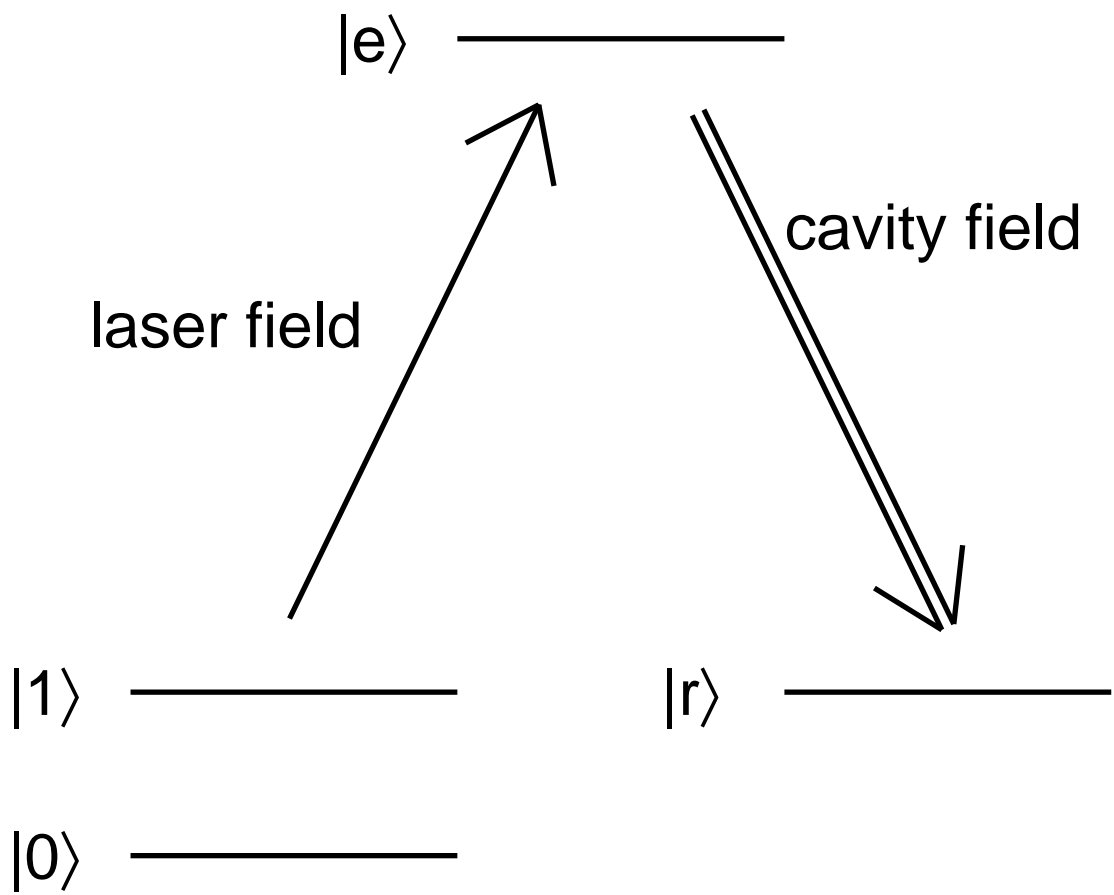
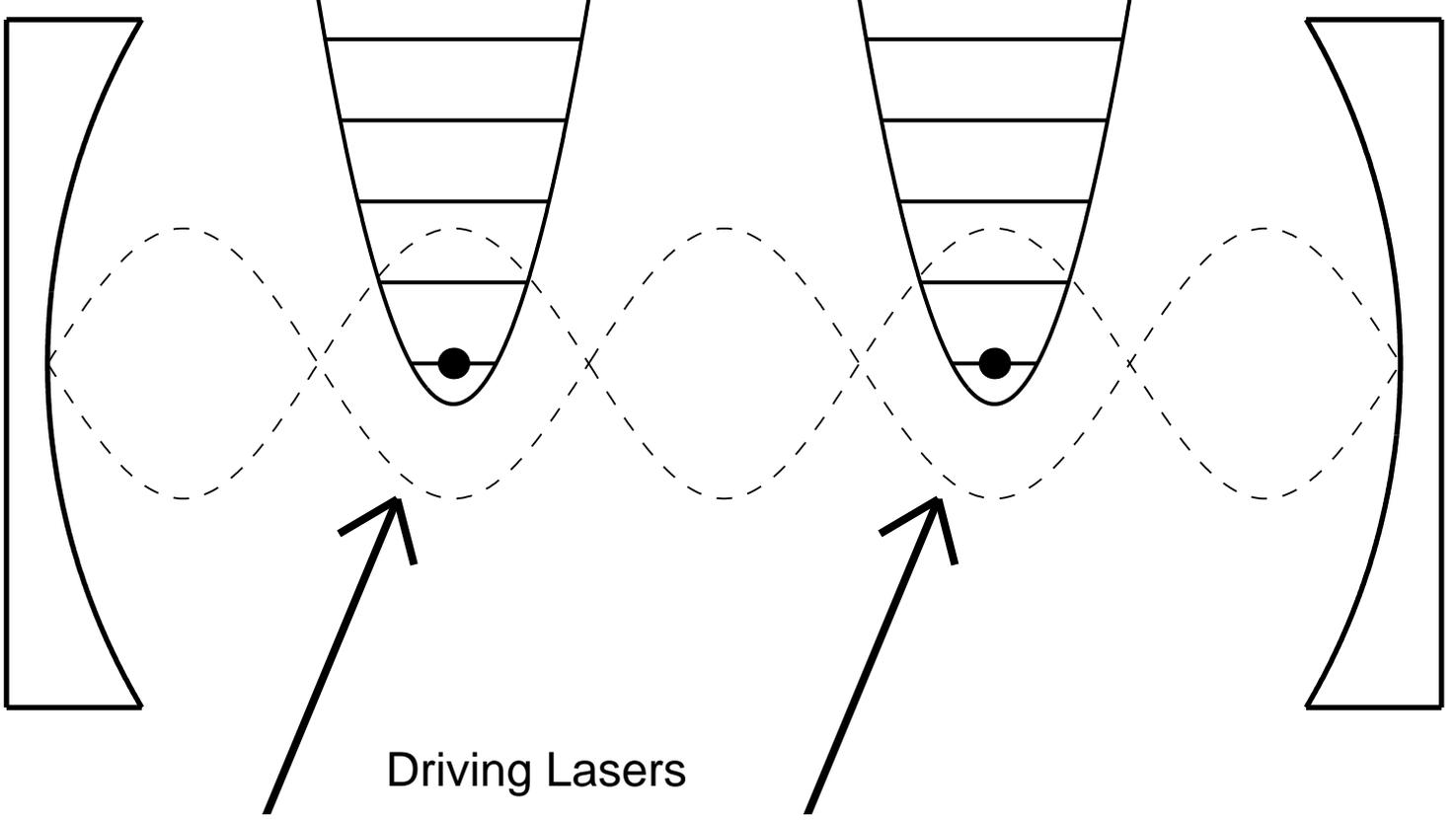
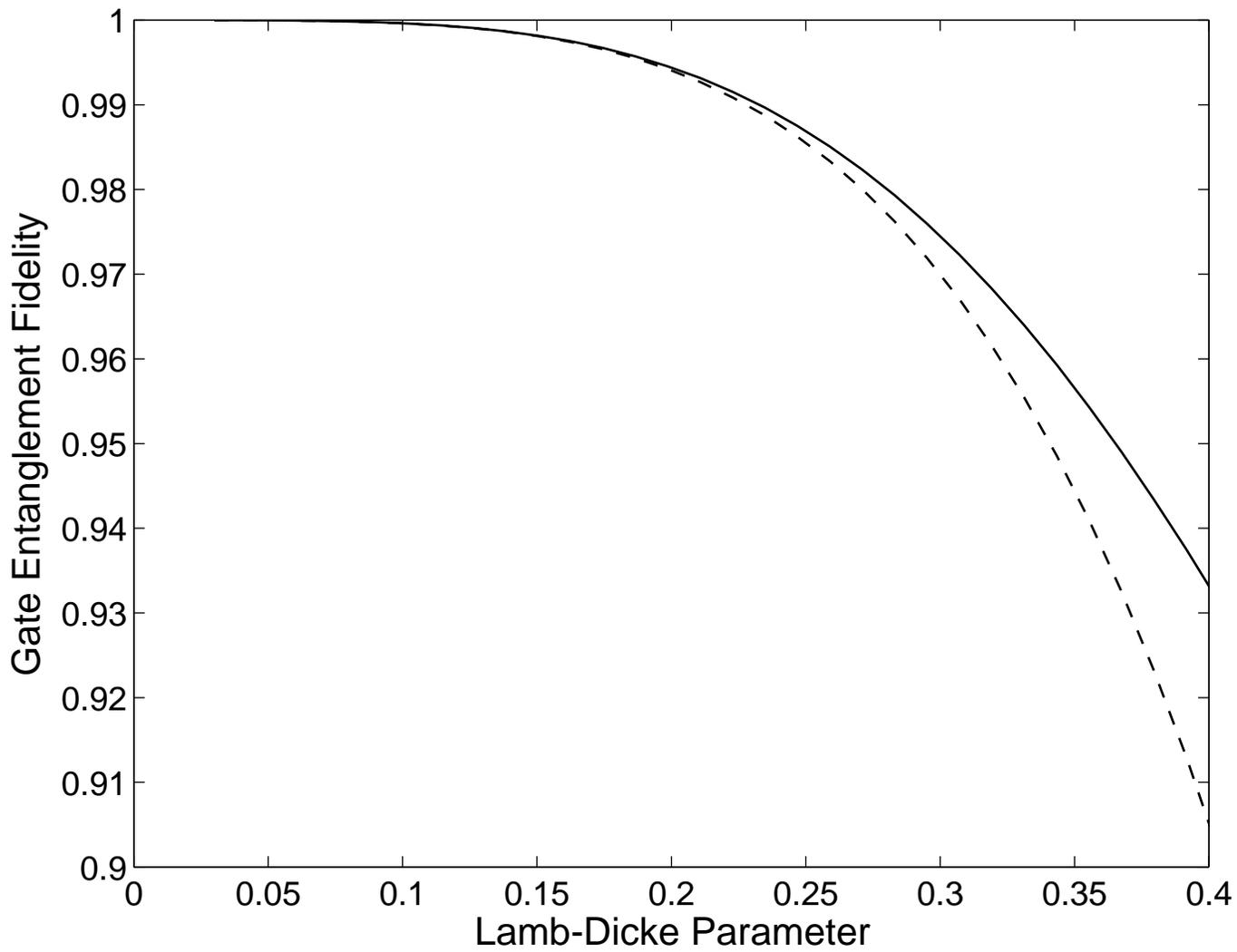
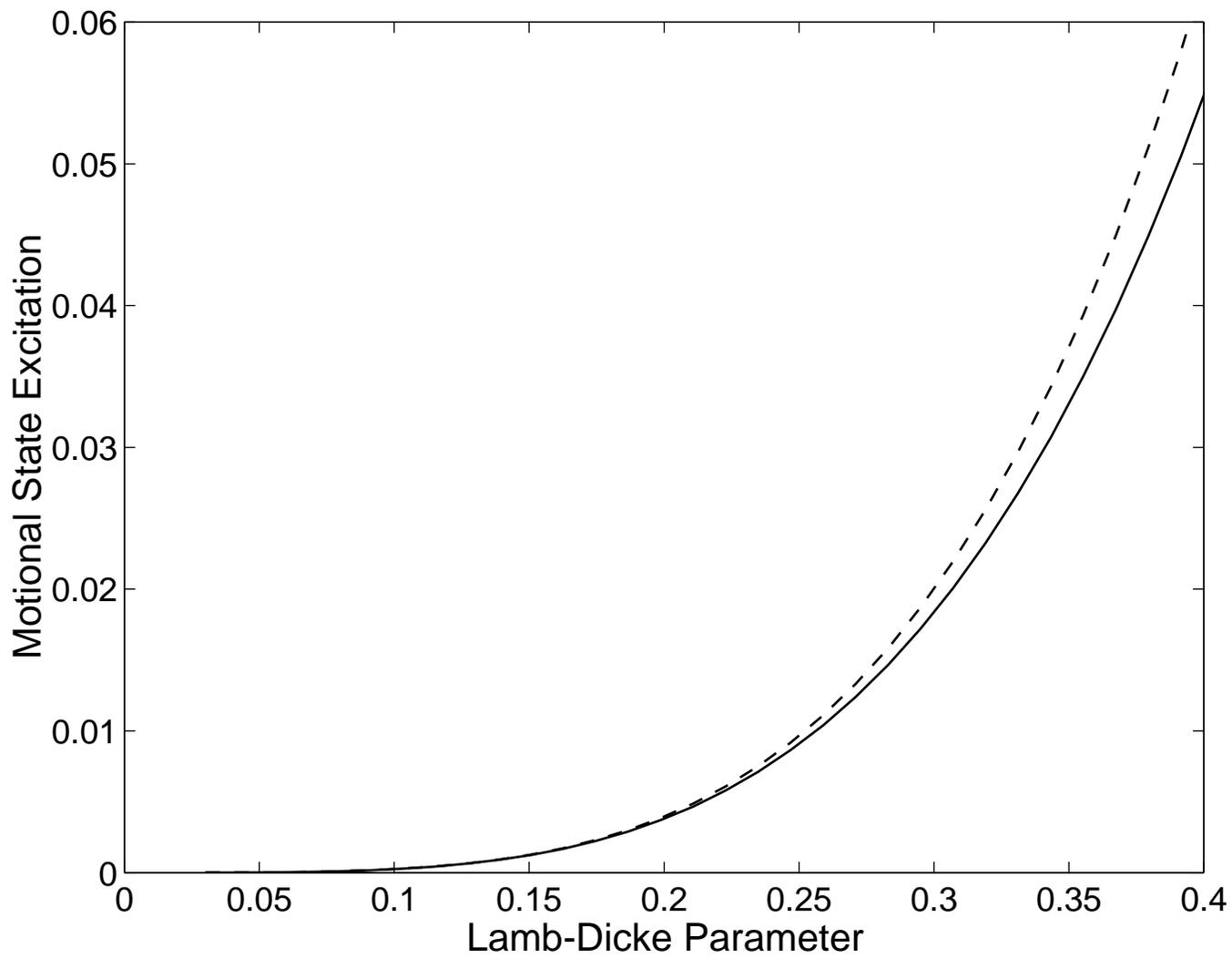
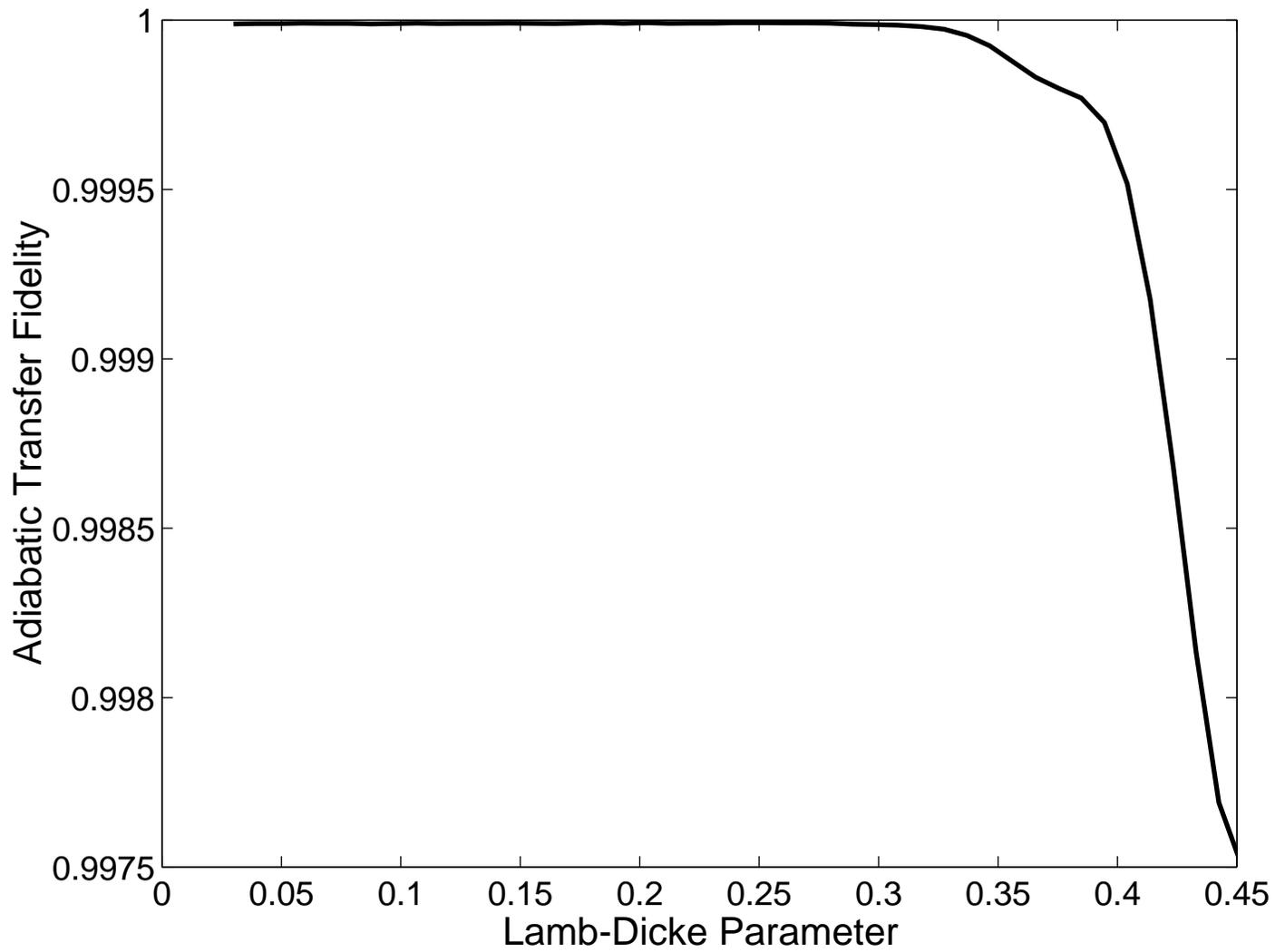
\end{document}